# Observations
# Concerning the probability of the existence of annihilators for balanced boolean functions

Hatem Y. Najdi

High Institute for Applied Science and Technology (HIAST), Damascus, Syria

najdi@hiast.edu.sy

**Abstract**: LFSR-based stream ciphers with nonlinear filters or combiners are susceptible to algebraic attacks using linearization methods to solve an overdefined system of nonlinear equations. And this process is greatly enhanced if the filtering or combining function has a low degree annihilator. To prevent such an attack, one would choose the parameters of that function so that the degree of its annihilator becomes large enough. As computing power is continuously increasing, a choice that seems secure today, becomes insecure tomorrow. Therefore, a tool is needed to estimate the probability of the existence of annihilators for balanced boolean functions with parameters that are beyond the current computing power. Based on experimental and calculational observations, we give in this paper an almost exact estimate of that probability, which represent a great improvement over the upper bound previously known.

**Keywords** : Annihilator, Stream ciphers, Algebraic attacks, Boolean functions, Algebraic degree,.

## 1 Introduction

LFSR stream ciphers with combining nonlinear functions are very efficient in terms of their speed of generating the keystream and their ability to mathematical analysis with respect of controlling the period of their generated sequences. Although these ciphers are nonlinear, they are not immune against algebraic attacks. Indeed, there exists efficient algorithms for linearization and solving nonlinear sets of equations [6], when the degree of the combining function, which is ultimately determined by the number of its inputs, is not too large. Moreover, there also exist algorithms that can find low degree annihilators of such functions so that they can be used to simplify the linearization process, and consequently, recover the encryption key [1, 5]. To escape such an attack, one would increase $n$, the number of the inputs to the combining function, which leads to a larger degree $d$ of its annihilators, and therefore increases its algebraic immunity. The question now is how big $n$ and $d$ should be in order to have a secure system. Using the annihilator's finding algorithms, one can check whether a given combining function has an annihilator, but this process is not sufficient to make a concrete decision about the security of a system. A choice of $n$ and $d$, that is secure today according to the mentioned checking process, will be insecure tomorrow due to the increasingly growing computing power. And choosing large number of inputs to combining function arbitrarily is not practical. True, the prices of bits are decreasing fast, but there are practical limits that can not be exceeded, especially because the memory requirements of the system grow exponentially with that number. In [5], an upper bound of the probability of the existence of balanced boolean functions annihilators has been derived, so one can predict at what range that number should be. Although that upper bound is good for practical purposes, it suffers from a fatal drawback. In fact, the value of the bound is very overestimated, giving probability values "greater" than 1 for certain values of $n$ an $d$. And in [2], a better upper bound is derived. In what follows, we present an almost exact formula for the probability of the existence of annihilators of a combining nonlinear boolean function, based on some empirical and calculational observations, but without a regirous mathematical proof. Our interest here is focused on balanced boolean functions, because balanced functions are a requirement in cipher systems.

The paper is organized as follows. In section 2 some definitions are given. In section 3, the algorithms of finding annihilators are quoted from [5]. In section 4, some experimental observations concerning the behavior of annihilators are described. Finally, in section 5, an analysis of the probability of the existence of annihilators for balanced boolean functions is made, following the approach of [5].

## 2 Definitions

We present some definitions that are necessary for the rest of the paper.

**Definition 1** [5]: Let $f : \{0, 1\}^n \to \{0, 1\}$ be a boolean function of n inputs. We call $g : \{0, 1\}^n \to \{0, 1\}$, $g \neq 0$, an annihilator of $f$ if $f \cdot g = 0$.

**Definition 2**: Let $x = [x_1, x_2, \ldots, x_n]$, $x_i \in \{0,1\}$, represent the n inputs to f. Then $f(x)$ can be represented by the Algebraic Normal Form ANF:

$$f(x) = a_0 \oplus a_1 x_1 \oplus \cdots \oplus a_n x_n \oplus \cdots \oplus a_{12} x_1 x_2 \oplus \cdots \oplus a_{1n} x_1 x_n \oplus \cdots \oplus a_{12..n} x_1 x_2 .. x_n$$

where $\oplus$ is the addition operator over the Galois field $\mathbf{F}_2$, and $a_i \in \mathbf{F}_2$.

In the ANF, the coefficient $a_0$ is the $0^{th}$ order coefficient, and the coefficients $a_1, \cdots, a_n$ are the $1^{st}$ order coefficients, and the coefficients $a_{12}, \cdots, a_{n-1n}$ are the $2^{nd}$ order coefficients, and so on. The coefficient $a_{12..n}$ is the $n^{th}$ order coefficient. The degree of $f$ is the number of variables $\{x_j\}$ associated with the highest order non-zero coefficient. The number of all ANF coefficients is $\sum_{i=0}^{n} \binom{n}{i} = 2^n$.

**Corollary 1**. *From definition 2, it follows that:*

*a) If $g(x)$ is of degree d, then the maximum number of terms in its ANF equals $s = \sum_{i=0}^{d} \binom{n}{i}$, and the total number of all functions g of degree d is $2^s$.*

*b) if $d < \lfloor n/2 \rfloor$, then $s < 2^{n-1}$.*

*c) if n is odd and $d = \lfloor n/2 \rfloor$, then $s = 2^{n-1}$.*

*d) if n is even and $d = n/2$, then $s > 2^{n-1}$.*

*e) if $d > n/2$, then $s > 2^{n-1}$.*

**Definition 3**. *The Hamming weight of f, or its weight $wt(f)$, is equal to the number of ones in its truth table. f is called balanced when the number of ones in its truth table equals the number of zeros, i. e. when $wt(f) = 2^{n-1}$.*

## 3 Finding the annihilators of $f$

Let $g(x) \neq 0$ be of degree $d \leq \lceil n/2 \rceil$. A necessary and sufficient condition for g to be an annihilator of f, i.e. for $f \cdot g = 0$, is that g vanishes for all arguments x for which $f(x) = 1$. This leads to the following algorithm:

**Algorithm 1** (quoted from [5]):

- *Substitute all N arguments x with $f(x) = 1$ in the ANF of a general boolean function g(x) of degree d. This gives a system of N linear equations for the coefficients of g(x).*

- *Solve this system of linear equations.*

- *If there is no ( nontrivial) solution, output no annihilator of degree d, else determine sets of coefficients for linearly independent annihilators.*

Of course, the number of unknowns in this system of linear equations is the number of the coefficients in the ANF of $g$, which equals $s$. Since we are interested here in balanced functions only, the number of equations that we can form, i.e. the value of $N$, will be $2^{n-1}$. Note that this system of linear equations is homogeneous because $g(x) = 0$ when $f(x) = 1$.

Algorithm 1 is the basic algorithm for finding the annihilators of a boolean functions $f$. In [5], there is another, more computationally efficient algorithm, but in essence, it is the same as algorithm 1, although it may sometimes give wrong results.

**Algorithm 2** (quoted from [5]):

1. Let weight $w = 1$.
2. For all x of weight w with $f(x) = 1$ substitute x in $g(x) = 0$ to derive a linear equation in the coefficients of g, with a single coefficient of weight w. Use this equation to express this coefficient iteratively by coefficients of lower weight.
3. If $w < d$, increment w by 1 and go to step 2.
4. Choose random arguments x of arbitrary weight such that $f(x) = 1$ and substitute in $g(x) = 0$, until there are the same number of equations as unknowns.
5. Solve the linear system. If there is no solution, output no annihilator of degree d.

Algorithm 2 is aimed at showing that f has no annihilator of given degree d. However, if the system turns out to be solvable, one may try another set of arguments x in step 5. If the new system is again solvable, one checks whether the solutions found are consistent. In case the number of variables n of f is not too large, one may directly verify whether one has found an annihilator, by formally expanding $f(x) \cdot g(x)$ and by checking whether the result is identically 0.

Note that if algorithm 2 is used to find an annihilator of a balanced function $f$, then all the $2^{n-1}$ equations must be involved in the process of solving the system. If one takes, for example, only $m$ equations such that $s < m < 2^{n-1}$, then the result will be for a function $f$ whose weight is $m$, not $2^{n-1}$. But there is nothing in algorithm 2 that ensures that it takes all those $2^{n-1}$ equations into account. The double checking process of the solutions may partially compensate for that, but it is not sufficient on its own unless $f(x) \cdot g(x) = 0$ is verified directly.

## 4 Empirical Observations

In [5], an upper bound is given for the probability of the existence of annihilators for balanced boolean functions by theorem 4. This upper bound suffers from being overestimated, and worse, it gives in certain cases probability values "greater" than 1, whereas it is much less than 1, e.g. when $n = 10$ and $d = 4$ [5].

Searching for annihilators for $n = 10$ and $d = 4$, by repetitively choosing random balanced functions $f$ and applying algorithm 1, shows that this probability must be very small. Therefore there must be something wrong in theorem 4 of [5].

During the search for annihilators, the following observations have been made:

1. For $d < \lfloor n/2 \rfloor$, it was noticed that the weights of the annihilators equal, almost always, the minimum weight $2^{n-d}$, and very rarely, $3 \times 2^{n-d-1}$, which is the weight next to the minimum one. This led to the conjecture that functions g with other weights have a very low probability of being annihilators of balanced functions f when $d < \lfloor n/2 \rfloor$. To see this more clearly, and due to the limited resources of computing, we searched for annihilators of randomly selected functions f with weights $wt(f)$ increasing from s towards $2^{n-1}$, computed the average weights of the annihilators $\overline{wt(g)}$ for $d < \lfloor n/2 \rfloor$, and plotted $\overline{wt(g)}/2^{n-d}$ against $wt(f) - s$ as shown in Fig. 1. Table 1 shows the curves parameters.

As can be seen from Fig. 1, the average value of the weights of the annihilators $g$ decreases as the weights of the functions $f$ increase. And well before the weights $wt(f)$ reach the value $2^{n-1}$, i.e. before $f$ becomes balanced, the average weights $\overline{wt(g)}$ approach the value $2^{n-d}$, confirming the conjecture that a balanced function $f$ with $n$ inputs mostly admits annihilators $g$ of degree $d$ whose weights are $2^{n-d}$. For example, for $n = 8$ and $d = 3$, and amongst 1712 annihilators, there has been 1710 of weight 32, and 2 of weight 48, giving an average weight of 32.01869. And for $n = 6$ and $d = 2$, and amongst 4456 annihilators, there has been 4441 of weight 16, and 15 of weight 24, giving an average weight of 16.0269.

2. Also for $d < \lfloor n/2 \rfloor$, and during the search for annihilators of randomly selected functions $f$ with weights $wt(f)$ increasing from $s$ towards $2^{n-1}$, we computed $\overline{r}$, the average value of the ranks of the coefficient matrices of the homogeneous equations when there are nontrivial solutions, i.e. when there are annihilators of $f$, and plotted $\overline{r}/(s-1)$ against $wt(f) - s$ in Fig. 2 for all pairs of $n$ and $d$ listed in table 1. As can be seen from Fig. 2, the average rank $\overline{r}$ approaches the number of unknowns $s$ minus 1 as the weight of $f$ increases towards $2^{n-1}$, which suggests that a balanced boolean function $f$ admits only one annihilator if its degree .

3. For $n$ odd and $d = \lfloor n/2 \rfloor$, we found that many annihilators of different weights annihilate balanced boolean functions $f$.

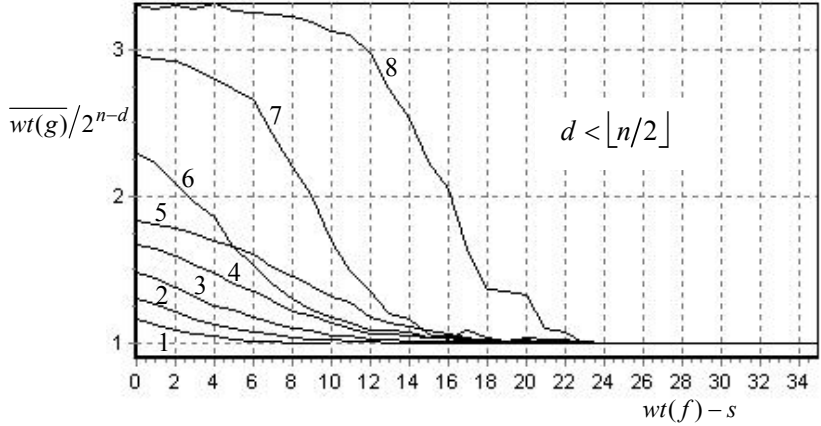

**Fig. 1**

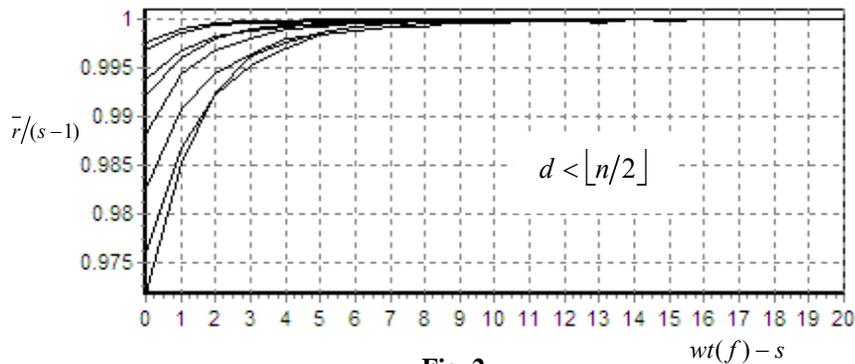

**Fig. 2**

| Curve no. | 1 | 2 | 3 | 4 | 5 | 6 | 7 | 8 |
|---|---|---|---|---|---|---|---|---|
| $n$ | 6 | 7 | 8 | 9 | 10 | 8 | 9 | 10 |
| $d$ | 2 | 2 | 2 | 2 | 2 | 3 | 3 | 3 |
| $2^{n-1} - s$ | 10 | 35 | 91 | 210 | 456 | 35 | 126 | 336 |

**Table 1**

## 5 The probability of the existence of annihilators for random balanced boolean functions $f$ of $n$ variables

Before proceeding with the discussion of the probability, it is worth mentioning that algorithm 1 leads to the following two lemmas.

**Lemma 1.** *The $2^n$ linear homogenous equations, form which algorithm 1 selects the system of N equations, are reducible to exactly s linearly independent equation for any d such that $s \leq 2^n$.*

*Proof.* If $wt(f) = 2^n$, algorithm 1 selects $N = 2^n$ equations. But when $wt(f) = 2^n$, $f(x) = 1$ for all $x \in \{0,1\}^n$. Consequently $g = 0$, and therefore there is no nontrivial solution of the system of the selected linear equations. Thus the s equations resulting from the reduction of the $2^n$ equations are linearly independent. □

**Lemma 2.** *Any set of $N < 2^n$ equations selected by algorithm 1 are reducible to less then s linearly independent equation with a probability greater then 0 for any d such that $s < 2^n$.*

*Proof.* If $wt(f) < 2^n$, algorithm 1 selects $N = wt(f) < 2^n$ equations. But when $wt(f) < 2^n$, then $f(x) = 0$ for some $x \in \{0,1\}^n$, giving rise to the possibility that $g \neq 0$, and therefore to the possibility of there being some nontrivial solutions of the system of selected equations. □

Now we will distinguish between the four cases of corollary 1: $d < \lfloor n/2 \rfloor$, $n$ odd and $d = \lfloor n/2 \rfloor$, $n$ even and $d = n/2$, and $d > n/2$.

### 5.1 $d < \lfloor n/2 \rfloor$

Since we are interested only in balanced boolean functions $f$, then the weight of $f$ is given by $wt(f) = 2^{n-1}$. This means that the weight $wt(g)$ of the annihilator $g$, is less than or equal to the weight of $f$, i.e $wt(g) \leq 2^{n-1}$. Now, for every such function $g$, the number of balanced functions $f$ such that the set of arguments $x$ for which $g = 1$ is included in the set of arguments for which $f = 0$ is $N_f = \binom{2^n - wt(g)}{2^{n-1} - wt(g)}$. It should be clear that $N_f$ is the number of balanced functions $f$ admitting an annihilator $g$ of weight $wt(g)$.

Let $A_w$ be the number of functions $g$ whose weights equal $w$. And let $\{f\}_i$ denote all balanced functions $f$ that admit annihilator $g_i$. Now, conjecturing that observation 2 is true, then a balanced function $f$ can admit only one annihilator of degree $d < \lfloor n/2 \rfloor$. This leads to the assumption that the $\{f\}_i$ are disjoint, and consequently, the total number of balanced functions $f$ that can be annihilated by these functions $g$ is equal to $A_w \cdot N_f$. Since the total number of balanced functions $f$ is $\binom{2^n}{2^{n-1}}$, the probability of there being an annihilator of weight $w$ for a balanced function $f$ is given by:

$$P(w) = A_w \frac{\binom{2^n - w}{2^{n-1} - w}}{\binom{2^n}{2^{n-1}}} \quad (1)$$

Consequently, the probability of the existence of annihilators for balanced boolean functions $f$ can be expressed as being equal to the sum of $A_w \dfrac{\binom{2^n - w}{2^{n-1} - w}}{\binom{2^n}{2^{n-1}}}$ over all possible values of the weights $w$ that are not greater than $2^{n-1}$. Namely:

$$P_{an} = \sum_{w(n,d) \leq 2^{n-1}} A_w \frac{\binom{2^n - w}{2^{n-1} - w}}{\binom{2^n}{2^{n-1}}} \quad (2)$$

In fact, $A_w$ is the number of codewords of weight $w$ in the Reed-Muller code $R(d,n)$, and the $A_w$ have been enumerated for $2^{n-d} \leq w < 2^{n-d+1}$ in [3,4], where $2^{n-d}$ is the minimum weight of $g$.

Now define $\alpha_1$, $\alpha_2$ and $\alpha_3$ such that:

$$\alpha_1 = A_{2^{n-d}} \frac{\binom{2^n - 2^{n-d}}{2^{n-1} - 2^{n-d}}}{\binom{2^n}{2^{n-1}}} \qquad \alpha_2 = \sum_{w > 2^{n-d}}^{w < 2^{n-d+1}} A_w \frac{\binom{2^n - w}{2^{n-1} - w}}{\binom{2^n}{2^{n-1}}} \qquad \alpha_3 = \sum_{w = 2^{n-d+1}}^{w = 2^{n-1}} A_w \frac{\binom{2^n - w}{2^{n-1} - w}}{\binom{2^n}{2^{n-1}}}$$

Then equation (2) becomes:

$$P_{an} = \alpha_1 + \alpha_2 + \alpha_3 \quad (3)$$

Since the enumerators for $2^{n-d} \leq w < 2^{n-d+1}$ are known, the evaluation of $\alpha_1$ and $\alpha_2$ is straightforward. For $\alpha_3$, there are to distinct cases: $d = 2$ and $d > 2$. When $d = 2$, then $2^{n-d+1} = 2^{n-1}$, and $\alpha_3$ becomes:

$$\alpha_3^{d=2} = A_{2^{n-1}} \binom{2^n}{2^{n-1}}^{-1} \quad (4)$$

where:

$$A_{2^{n-1}} = 2^s - 2 \sum_{w=2^{n-2}}^{w < 2^{n-1}} A_w$$

This is because the $A_w$ are symmetric about $w = 2^{n-1}$ and $\sum_{w=0}^{2^n} A_w = 2^s$. Consequently, $\alpha_3^{d=2}$ can also be directly evaluated. Unfortunately, when $d > 2$, things are not so obvious.

It has been shown in [8] that if the codeword in the $d^{\text{th}}$-order Reed Muller code of length $2^n$ has weight $w$ less than twice the minimum weight $2^{n-d}$, then $w$ is of the form:

$$w = 2^{n-d+1} - 2^{n-d+1-\mu} \qquad \text{for some positive } \mu \quad (5)$$

For $\mu = 1$, equation (5) gives $w = 2^{n-d}$, which has been enumerated in [4] as:

$$A_{2^{n-d}} = 2^d \prod_{i=0}^{n-d-1} \frac{2^{n-i} - 1}{2^{n-d-i} - 1} \tag{6}$$

And for $\mu \geq 2$, the enumerators $A_w$ are given by theorem 2 of [3] for $2^{n-d} < w < 2^{n-d+1}$.

Table 4 shows the values of $\alpha_1$ and $\alpha_2$ for several values of $n$ and $d$ satisfying $d < \lfloor n/2 \rfloor$. It is clear from the table that $\alpha_2 \ll \alpha_1$. And this is our calculational observation. Moreover, a closer inspection of $\alpha_2$ reveals that its terms decrease very rapidly with increasing $w$. This fact, together with observation 1, leads us to conjecture again that the terms of $\alpha_3$ are also vanishingly negligible, making $\alpha_1$, i.e. the first term of equation (2), the dominant one. This allows by all means expressing the final result that gives the probability of the existence of an annihilator $g$ of degree $d$ for a balanced function $f$ with $n$ inputs, when $d < \lfloor n/2 \rfloor$, in the following form with a very good accuracy:

$$P_{an}^{d<\lfloor n/2 \rfloor} \approx \alpha_1 = A_{2^{n-d}} \frac{\binom{2^n - 2^{n-d}}{2^{n-1} - 2^{n-d}}}{\binom{2^n}{2^{n-1}}} \tag{7}$$

Note that for $n = 10$ and $d = 4$, $P_{an}$ is very small as expected. Remember that theorem 4 in [5] gives a probability greater than 1 for this case. For $n = 10$ and $d = 3$, the obtained results here show that the value obtained in [5] is very overestimated.

Table 5 shows some results that have been obtained experimentally by applying algorithm 1 of [5] on randomly selected balanced functions $f$ for the case $d < \lfloor n/2 \rfloor$. For $n = 6$ and $d = 2$, the experimental probability $P_{ex}$ is very close to $P_{an}$, and that is due to the relatively large number of tested functions $f$. For $n = 7$ and $d = 2$, and $n = 8$ and $d = 3$, the number of tested functions was relatively small compared with what the law of large numbers requires. Nevertheless, the values of $P_{ex}$ is not very far from the values of $P_{an}$, and this assures that formula (7) is correct, which means that our conjectures made above are reasonable.

For a given $d$, it is also clear from table 4 that as $n$ increases, $P_{an}^{d<\lfloor n/2 \rfloor}$ decreases. This is due to the following lemma.

**Lemma 3.** $P_{an}^{d<\lfloor n/2 \rfloor} \to 0$ when $n \to \infty$.

*Proof. Replacing the factors of the product* $\prod_{i=0}^{n-d-1} \frac{2^{n-i} - 1}{2^{n-d-i} - 1}$ *in equation (6) by the biggest one gives:*

$A_{2^{n-d}} < 2^d \left(2^{d+1} - 1\right)^{n-d}$. *And noting that* $\frac{\binom{2^n - 2^{n-d}}{2^{n-1} - 2^{n-d}}}{\binom{2^n}{2^{n-1}}} \leq 2^{-2^{n-d}}$, *then formula (7) becomes:*

$$P_{an}^{d<\lfloor n/2 \rfloor} < 2^d \left(2^{d+1} - 1\right)^{n-d} 2^{-2^{n-d}} < 2^{n(d+1) - d^2 - 2^{n-d}} \tag{8}$$

*Inequality (8) implies that* $P_{an}^{d<\lfloor n/2 \rfloor} \to 0$ *when* $n \to \infty$. □

Note that lemma 3 shows that theorem 3 of [5] is rather conservative, where $d$ is upper bounded by

$0.22 n$.

| $n$ | $d$ | $\alpha_1$ | $\alpha_2$ | $n$ | $d$ | $\alpha_1$ | $\alpha_2$ |
|---|---|---|---|---|---|---|---|
| 6 | 2 | 3.20E-3 | 1.23E-5 | 14 | 2 | 1E-1527 | 5E-2692 |
| 7 | 2 | 1.32E-8 | 6E-15 | 14 | 3 | 1E-669 | 3E-1062 |
| 8 | 2 | 5E-20 | 7E-35 | 14 | 4 | 5E-310 | 1E-476 |
| 8 | 3 | 1.97E-5 | 4.3E-8 | 14 | 5 | 7E-143 | 3E-217 |
| 9 | 2 | 2E-43 | 6E-76 | 14 | 6 | 7E-62 | 4E-94 |
| 9 | 3 | 4E-15 | 2E-23 | 18 | 5 | 1E-2502 | 1E-3799 |
| 10 | 2 | 1E-90 | 3E-159 | 18 | 6 | 8E-1224 | 2E-1848 |
| 10 | 3 | 2E-35 | 1E-55 | 18 | 7 | 6E-595 | 8E-898 |
| 10 | 4 | 6E-12 | 6E-18 | 18 | 8 | 8E-283 | 3E-428 |
| 11 | 2 | 4E-186 | 3E-327 | 19 | 6 | 4E-2469 | 1E-3727 |
| 11 | 3 | 4E-77 | 1E-121 | 19 | 7 | 8E-1213 | 9E-1828 |
| 11 | 4 | 6E-31 | 3E-47 | 19 | 8 | 2E-589 | 1E-889 |
| 12 | 2 | 2E-377 | 2E-664 | 20 | 7 | 1E-2450 | 2E-3690 |
| 12 | 3 | 3E-161 | 3E-255 | 20 | 8 | 2E-1205 | 3E-1815 |
| 12 | 4 | 4E-70 | 2E-107 | 20 | 9 | 5E-585 | 2E-883 |
| 12 | 5 | 1E-27 | 7E-42 | 21 | 8 | 2E-2439 | 1E-3669 |
| 13 | 2 | 1E-760 | 7E-1340 | 21 | 9 | 1E-1199 | 8E-1807 |
| 13 | 3 | 2E-330 | 1E-523 | 22 | 9 | 9E-2432 | 6E-3657 |
| 13 | 4 | 1E-149 | 6E-230 | 22 | 10 | 6E-1195 | 2E-1800 |
| 13 | 5 | 2E-65 | 1E-99 | 23 | 10 | 8E-2426 | 3E-3648 |

**Table 4**

| $n$ | $d$ | $P_{ex}$ | $P_{an}$ |
|---|---|---|---|
| 6 | 2 | 3.22E-3 | 3.20E-3 |
| 7 | 2 | 1.195 E-8 | 1.32E-8 |
| 8 | 3 | 2.092 E-5 | 1.97E-5 |

**Table 5**

## 5.2 $n$ odd and $d = \lfloor n/2 \rfloor$

By corollary 1-c, $s = 2^{n-1}$, i.e. the number of available equations for algorithm 1 is equal to the number of unknowns. But by lemma 1 and lemma 2, these equations are not necessarily all linearly independent, and therefore the possibility of there being some non trivial solutions exists. On the other hand, removing just one equation from these equations leads always to non trivial solutions, i.e. the system of equations is on the verge of avalanche. This suggests that the probability of the existence of annihilators must be high.

According to observation 3, the annihilators can have many different weights whose enumerators are unknown, and therefore no further concrete statement can be made about the probability of annihilators. However, experiments suggest that:

$$P_{ex}^{d=\lfloor n/2 \rfloor} > 0.7 \quad (10)$$

as shown in table 6.

| $n$ | $d$ | $P_{ex}^{d=\lfloor n/2 \rfloor}$ |
|---|---|---|
| 7 | 3 | 0.842 |
| 9 | 4 | 0.717 |
| 11 | 5 | 0.712 |
| 13 | 6 | 0.70 |
| 15 | 7 | 0.72 |

**Table 6**

### 5.3 $n$ even and $d = n/2$

By corollary 1-d, $s > 2^{n-1}$, i.e. the number of available equations is less than the number of unknowns. This makes the system of linear equations in algorithm 1 underdefined, and therefore there is always non trivial solutions, giving:

$$P_{an}^{d=n/2} = 1 \quad (9)$$

### 5.4 $d > n/2$

By corollary 1-e, $s > 2^{n-1}$, and the result here is identical to that given in 5.2.

## 6 Conclusions

The main result of this work was formula (7). In fact, we have made conjectures, based on empirical and calculational observations, that a balanced boolean function with $n$ inputs admits annihilators of degree $d < \lfloor n/2 \rfloor$ whose weights are equal to $2^{n-d}$ almost always, and to $3 \cdot 2^{n-d-1}$ very rarely. These observation enabled the construction of formula (7) that gives the probability of the existence of annihilators for balanced boolean functions very accurately, instead of the loose upper bound known before, for the case $d < \lfloor n/2 \rfloor$.

Formula (7) is the most important in this work, because of the difficulty, even the impossibility, of determining the probability under study by other means. The cases of subsections 5-2 to 5-4 have been included here for completeness.

**Acknowledgement**
The author would like to thank Omran Kouba for his helpful discussions and for reviewing the final version of this paper.